\RequirePackage{fix-cm}
\documentclass{article}                     
\usepackage{graphicx}
\graphicspath{ {./images/} }
\usepackage{xcolor}
\definecolor{sapphirecrayola}{rgb}{0.18,0.36,0.63}
\definecolor{aoenglish}{rgb}{0.0,0.5,0.0}
\usepackage{amssymb}
\usepackage{amsmath}
\usepackage{nccmath}
\usepackage{mathtools}

\usepackage{subfig}
\usepackage{placeins}
\usepackage{afterpage}
\edef\svtheparindent{\the\parindent}
\usepackage{parskip}
\parindent=\svtheparindent\relax
\usepackage{setspace}
\usepackage{enumitem}
\usepackage{comment}
\usepackage{hyperref}
\usepackage{graphicx,calc}
\newlength\myheight
\newlength\mydepth
\settototalheight\myheight{Xygp}
\usepackage{geometry}

\usepackage{natbib}
\setcitestyle{authoryear,open={(},close={)}}
\usepackage{todonotes}
\presetkeys{todonotes}{color=yellow!20}{}

\usepackage{graphicx}
\usepackage{booktabs}
\usepackage{caption}
\usepackage{float}

\usepackage{hyphenat}
\usepackage{hyperref}
\usepackage{pdfpages}

\begin{document}
\title{Blockchain mechanism and distributional characteristics of cryptos
\thanks{Financial support of the European Union’s Horizon 2020 research and innovation program “FIN- TECH: A Financial supervision and Technology compliance training programme” under the grant agreement No 825215 (Topic: ICT-35-2018, Type of action: CSA), the European Cooperation in Science \& Technology COST Action grant CA19130 - Fintech and Artificial Intelligence in Finance - Towards a transparent financial industry, the Deutsche Forschungsgemeinschaft’s IRTG 1792 grant, the Yushan Scholar Program of Taiwan and the Czech Science Foundation’s grant no. 19-28231X / CAS: XDA 23020303 are greatly acknowledged.}
}


\author{
Min-Bin Lin\thanks{International Research Training Group 1792, Humboldt-Universität zu Berlin, Spandauer Str. 1, 10178 Berlin, Germany. Email: min-bin.lin@hu-berlin.de} \and 
Kainat Khowaja\thanks{International Research Training Group 1792, Humboldt-Universität zu Berlin, Spandauer Str. 1, 10178 Berlin, Germany. Email: kainat.khowaja@hu-berlin.de} \and 
Cathy Yi-Hsuan Chen\thanks{Adam Smith Business School, University of Glasgow, United Kingdom; IRTG 1792 High Dimensional Non Stationary Time Series, Humboldt-Universit\"{a}t zu Berlin.  Email: cathyyi-hsuan.chen@glasgow.ac.uk } \and 
Wolfgang Karl Härdle\thanks{BRC Blockchain Research Center, Humboldt-Universität zu Berlin,  Berlin, Germany; Sim Kee Boon Institute, Singapore Management University, Singapore; WISE Wang Yanan Institute for Studies in Economics, Xiamen University, Xiamen, China; Dept. Information Science and Finance, National Chiao Tung University, Hsinchu, Taiwan, ROC; Dept. Mathematics and Physics, Charles University,  Prague, Czech Republic, Grants--DFG IRTG 1792 gratefully acknowledged. Email: haerdle@hu-berlin.de}
}
\date{}

\maketitle
\begin{abstract} 

We investigate the relationship between underlying blockchain mechanism of cryptocurrencies and its distributional characteristics. In addition to price, we emphasise on using actual block size and block time as the operational features of cryptos. We use distributional characteristics such as fourier power spectrum, moments, quantiles, global we optimums, as well as the measures for long term dependencies, risk and noise to summarise the information from crypto time series. With the hypothesis that the blockchain structure explains the distributional characteristics of cryptos, we use characteristic based spectral clustering to cluster the selected cryptos into five groups. We scrutinise these clusters and find that indeed, the clusters of cryptos share similar mechanism such as origin of fork, difficulty adjustment frequency, and the nature of block size. This paper provides crypto creators and users with a better understanding toward the connection between the blockchain protocol design and distributional characteristics of cryptos. 


\end{abstract}
\clearpage
\doublespacing

%
%
\label{secIntro}
\section{Introduction}








Cryptocurrency (crypto) is a digital asset designed to be as a \emph{medium of exchange} wherein individual coin ownership is recorded in a digital ledger or computerised database.
Its creation of monetary units and verification of fund transactions are secured using encryption techniques and distributed across several nodes (devices) on a peer-to-peer network. 
Such technology-enhanced and privacy-preserving features make it potentially different to other existing financial instruments and has attracted attention of many investors and researchers \citep{haerdle2020understanding}. 
Many studies have investigated the similarity between a pool of cryptocurrencies in order to classify the important features of digital currencies. 
For example, \cite{Blau2020} has concluded  that the top sixteen most active cryptocurrencies co-move with bitcoin.
Researchers have also focused on describing the price behaviour of cryptos using economic factors \citep{Pavel2016, sovbetov2018factors}. 
However, owing to the unique technology of cryptocurrencies, there still exists a gap between the creators of blockchain mechanism and users operating the financial market of the crytocurrencies and through this research, we aim to take a step towards mitigating that gap.  


We specialise our research on the following research questions. 
First, we characterise crypto behaviour using distributional characteristics of time series data.
Also, instead of using the prices alone, we use actual  block time and block size to incorporate the operational features of cryptos. 
Second, we hypothesise that the blockchain structure that the coin attaches plays a pivotal role in explaining the behaviour. More explicitly, we investigate the extent to which blockchain structure leads to explain the distributional characteristics. 
Using a characteristic based clustering coupled with spectral clustering technique, we group the selected cryptos into a number of clusters and stratify the mechanisms that make the coins within the particular cluster showing the same behaviour in price, actual block time, and actual block size, respectively.

When studying cryptocurrencies, many researchers only focus on crypto price and daily returns \citep{TRIMBORN2018107, Hou2020pricing}. 
While price is important when cryptos are used as a medium of payment, it is definitely not the only measure for evaluation of cryptocurrencies.
For example, many low price coins are highly traded and many coins that are not used as medium of payment have low prices, e.g., XPR and Dogecoin. 
Cryptos were introduced to serve various purposes and the purpose of the coin does matter. 
This makes it necessary to use other time series while studying crypto markets. 
In this research, we propose to use actual block size and actual block time alongside price.

Actual block size is the average actual size "usage" of a single block in data storage for one day. 
Since a block comprises of transaction data, it can represent the status of how a blockchain mechanism allocates transactions to a block.
We consider it a measure of scalability of the system. 
A well-functioning blockchain should be able to level the transaction arrivals. 
Transaction distribution within a day for any crypto needs such balancing because it affects miners rewards and hence the demand of the coin.  
An ideal block size would keep confirmation times from ballooning while keeping fees and security reasonable. 
Therefore, actual block size of cryptos can provide insight into the behaviour of cryptos. 

Actual block time, on the other hand, measures the consistency and performance of the system.
It is defined as the mean time required in minutes for each day to create the next block.
In other words, it is the average amount of time for the day a user has to wait, after broadcasting their transaction, to see this transaction appear on the blockchain. 
Think of crypto markets as a fast food franchise and miners as customers who have to wait a certain time to make the purchase. 
If the waiting time is shorter on certain days while on other instances, the customers have to wait much longer, there is a discrepancy in the system. 
Analogously, the time series of block time, which is the distribution of waiting time, can be seen as a service level of the whole system and it is necessary to maintain as the users' expectation or target block time set by the system depend on it.

The idea of investigating the underlying blockchain mechanism, a cornerstone of crypto technology, and its connection to the crypto behaviour is still in its infancy. One of the first endeavours in explaining this relationship was made by \cite{Li2018} who highlight that the the fundamental characteristics of cryptocurrencies (e.g., algorithm and proof type) have a vital role in differentiating the performance of cryptocurrencies. They develop a spectral clustering methodology to group cryptos in a dynamic fashion, but their research is limited in the exploitation of blockchain characteristics. With a similar spirit, \cite{Iwamura2019} start by claiming that high fluctuation is a reflection of the lack of flexibility in the Bitcoin supply schedule. 
They further strengthen their arguments by considering the predetermined algorithm of cryptos (specifically, the proof of work) to explain the volatility in cryptocurrency market. \cite{zimmerman2020blockchain} argue in their work that the higher congestion in blockchain technology leads to higher volatility in crypto prices. They claim that the limited settlement space in blockchain architecture makes users compete with one another, affecting the demand. In his model, the value of cryptos is governed by its demand, making the price sensitive to blockchain capacity.

These research results, albeit true, are limited to a particular set of cryptocurrency mechanism and do not thoroughly explain the dynamics of cryptocurrencies. Also, most of the papers only use price as a proxy of behaviour. We advance the previous findings by incorporating a rich set of underlying mechanisms and connecting them to multiple time series. We take a deep dive into eighteen cryptos with a variety of mechanisms- concluded in \cite{Garriga2020})- from a technical perspective to summarise their mechanism and algorithm designs using variables, such as consensus algorithm, type of hashing algorithm, difficulty adjustment frequency and so on. 

We investigate a relationship between underlying blockchain mechanism of cryptocurrencies and the distributional characteristics. Using the a characteristic-based clustering technique, we cluster the selected coins into a number of clusters and scrutinise the compositions of fundamental characteristics in each group. We observe that the clusters obtained from these time series indeed share common underlying mechanism. Through empirical evidence, we show that the cryptos forked from same origin and same consensus mechanism tend to become part of same clustering group. Furthermore, the clusters obtained by the time series of block time have same hashing algorithms and difficulty adjustment algorithms. Also, a similar nature (static or dynamic) of block size was observed within clusters obtained by the time series of actual block size. 
We conclude with empirical evidence that the crypto behaviour is actually linked with their blockchain protocol architectures.

The implications of this study are abundant. 
The creators of cryptocurrencies can manage the impact of blockchain underlying mechanisms on the corresponding distributional characteristics, in a consideration of adoption rate of invented coins. 
From the users' perspective, they can make an optimal decision in which coins should be adopted while concerning the price fluctuation. 

This paper proceeds as follows. 
\hyperref[secData]{Section 2} discusses data source and the underlying mechanisms of the cryptos.
\hyperref[secMeth]{Section 3} presents the methodology used for classifying characteristics of time series and clustering algorithm.
\hyperref[secRes]{Section 4} provides an illustration of analysis results.
\hyperref[secCol]{Section 5} concludes and provides several avenues for future research.

\label{secData}
\section{Data Source and Description}

According to CoinMarketCap (https://coinmarketcap.com), currently there are over 7,000 cryptocurrencies and their total market capitalisation has surpassed USD\$400 billion as of November 09, 2020.
Most of studies have focused on the mainstream coins (e.g., Bitcoin, Ethereum), and little has been investigated on the coins which have been introduced and featured with a diverse blockchain mechanisms and invented technologies. The work of \cite{guo2020bibliometrics} is one of exceptions. In this study, 18 cryptos with different set of blockchain mechanisms have been examined --Bitcoin, Bitcoin Cash, Bitcoin Gold, Bitcoin SV, Blackcoin, Dash, Dogecoin, Ethereum, Ethereum Classic, Feathercoin, Litecoin, Monero, Novacoin, Peercoin, Reddcoin, Vertcoin, XRP (Ripple), and Zcash. We explore an interplay between distributional characteristics of crpytos and blockchain mechanism. 
We discuss the key characteristics of blockchain mechanisms and the time series data in this section.

\subsection{Underlying Mechanism}
\begin{figure}[!htb]
    \centering
    \includegraphics[width=1\textwidth]{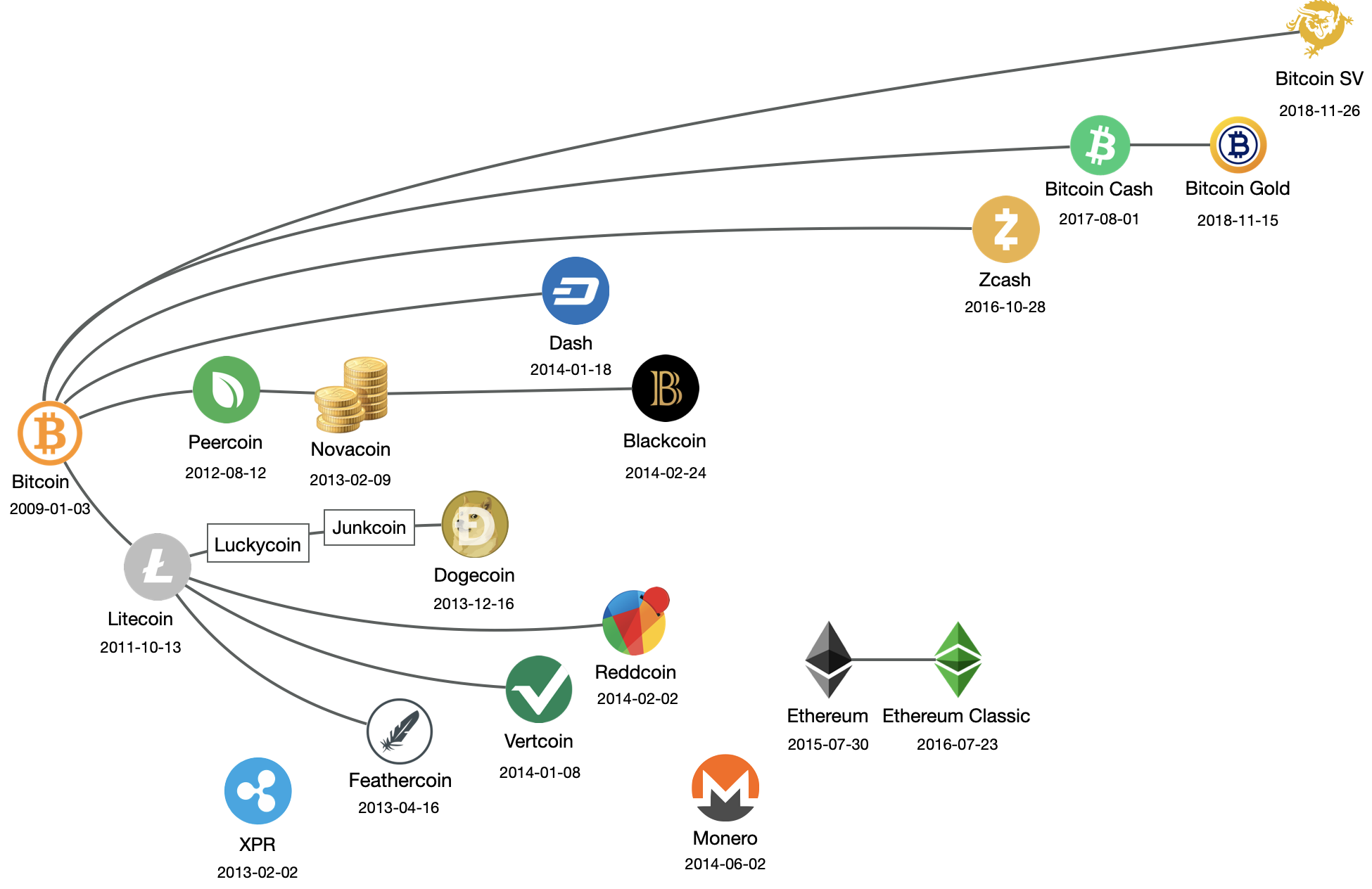}
    \caption{Blockchain software forks in cryptocurrency.}
    \label{fig:forks}
\end{figure}

Most of cryptos nowadays apply blockchain-based systems in which transactions are grouped into blocks and cryptographically interlinked to form a back-linked list of blocks containing transactions.
The transactions are validated using the nodes within the crypto peer-to-peer network through a majority consensus directed by algorithms instead of a central authority's approval.
In such an operation process, many algorithmic mechanisms are required to govern the performance and outcome of a crypto system.
Some key blockchain-based characteristics are discussed below:

\textbf{Fork:} It occurs as user base or developers conduct a fundamental or significant software change, see as in Figure~\ref{fig:forks}. There are two types of forks -- soft and hard forks. 
The former is an update to the protocol architecture and then all the nodes are enforced to follow in order to proceed with the operations of a crypto. 
The latter one creates a duplicate copy of the origin blockchain and modifies the copy to meet the desired quality (e.g., safety, scalability). 
In this case, a new crypto can be generated accordingly.
For example, Peercoin network facilitates an alternative consensus mechanism --proof-of-stake (PoS) to Bitcoin’s proof-of-work (PoW) system for reducing dependency on energy consumption from mining process \citep{king2012ppcoin}.


Going beyond a digital currency, Ethereum establishes an open-ended decentralised  platform for diverse applications such as decentralised applications (dapps) and smart contracts \citep{buterin2014next}.

\textbf{Consensus mechanism:} In order to establish an agreement on a specific subset of the candidate transactions, consensus mechanism provides a protocol for a large number of trust-less nodes in a decentralised blockchain network. 
For instance, PoW (Proof-of-Work, as adopted by e.g., Bitcoin, Litecoin) achieves consensus with a competition among miners on solving computational puzzles, which consume numerous computational resources; and PoS (Proof-of-Stake, as adopted by e.g., Peercoin, Blackcoin) randomly assigns a block creator (transaction validator) with probability proportional to their coins staked.

\textbf{Hashing algorithm:} It is a mathematical algorithm that encrypts a new transaction (or a new block) into a fixed length character string, known as hash value, and later interlinks this string with a given blockchain to ensure the security and immutability of a crypto. Various hashing algorithms are implemented in cryptos such as SHA-256, Scrypt and Equihash. These provide  different degree of complexity to blockchain operations.

\begin{figure}[!htb]
    \centering
    \includegraphics[width=1\textwidth]{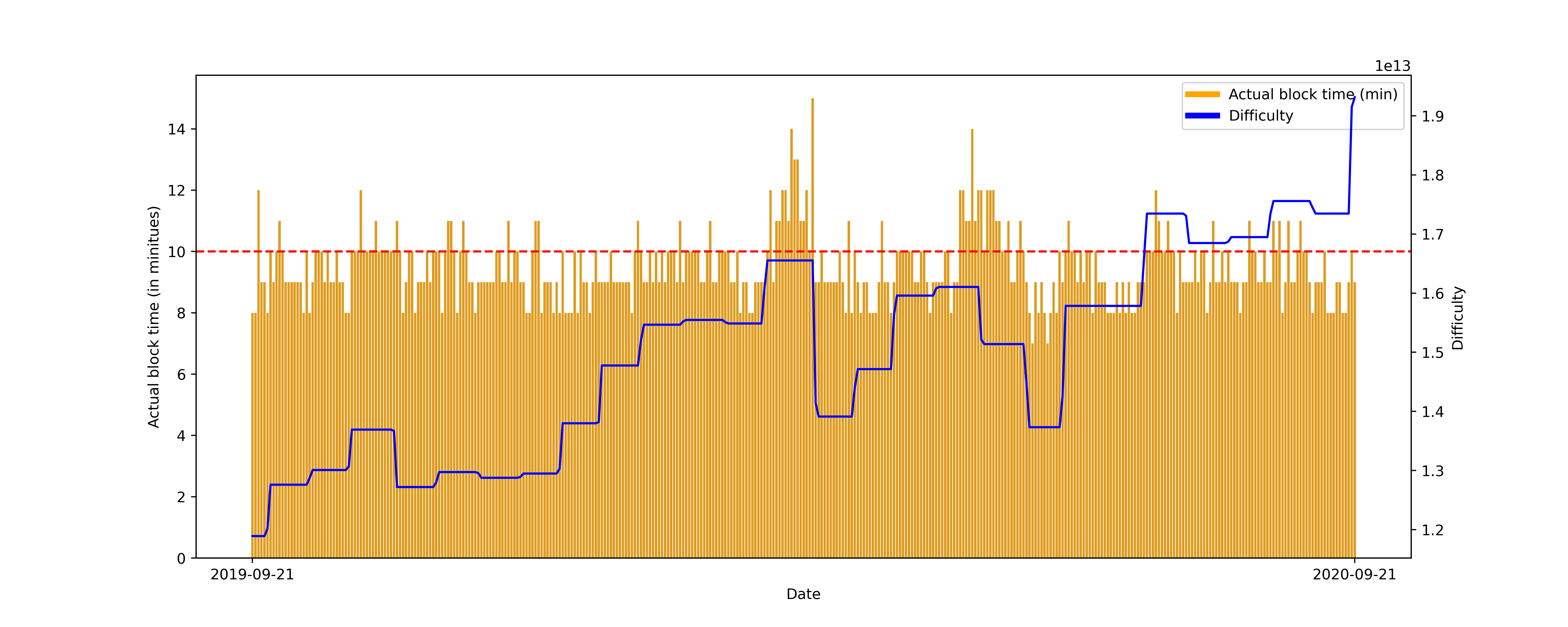}
    \caption{Bitcoin's difficulty adjustment toward actual block time.
    \protect \includegraphics[height=0.5cm]{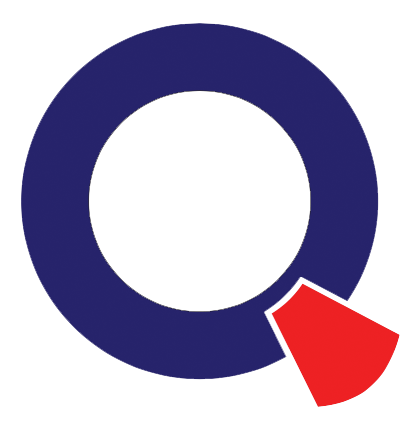} {\color{blue}\href{https://github.com/QuantLet/Blockchain\_mechanism/tree/main/Blockchain\_mechanism_plotting}{Blockchain\_mechanism\_plotting}}}
    \label{fig:diff_vs_time}
\end{figure}

\textbf{Difficulty adjustment algorithm:} It is an adaptive mechanism which periodically adjusts the difficulty toward hashrate to target an average time interval between blocks, known as target block time or target confirmation time. It regulates the creation rate of a block and maintains a certain amount of outputs of a blockchain. Such a mechanism is commonly seen in a PoW framework. An example from Bitcoin is shown in Figure~\ref{fig:diff_vs_time} where its difficulty adjustment algorithm, known as DAA, modifies the difficulty every 2016 blocks to meet target block time of 10 minutes. 

\subsection{Time Series Data}
The data applied in this paper are collected from Bitinfocharts which is available at https://bitinfocharts.com/. 
These time series are composed of data points observed daily from the genesis date of each crypto.
The lengths of these time series are thus varied coin by coin, but as explained in the section \ref{secMeth}, we continue to use the whole time series for each coin.

\begin{figure}[!htb]
    \centering
    \includegraphics[width=1\textwidth]{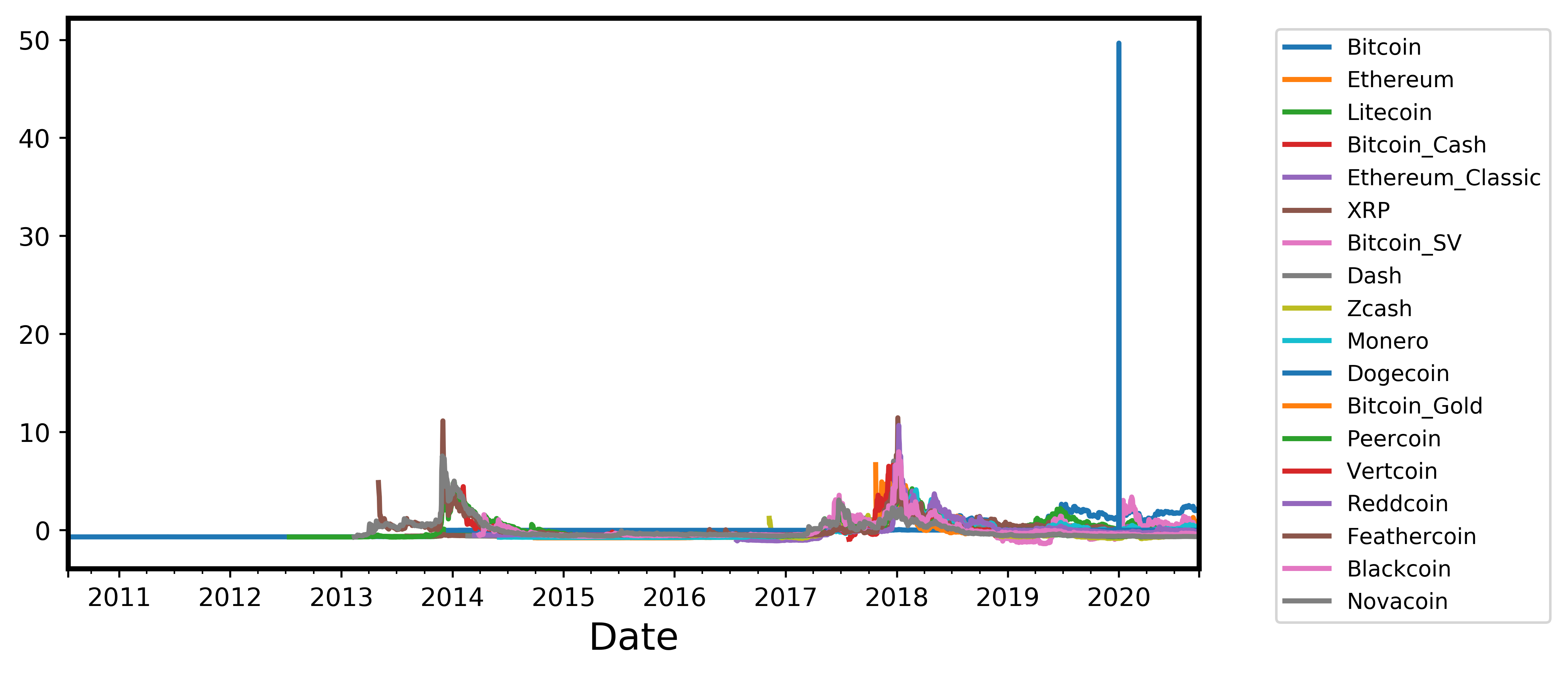}
    \caption{Time series of prices of the 18 cryptos
    \protect \includegraphics[height=0.5cm]{images/qletlogo_tr.png} {\color{blue}\href{https://github.com/QuantLet/Blockchain\_mechanism/tree/main/Blockchain\_mechanism_plotting}{Blockchain\_mechanism\_plotting}}}
    \label{fig:price}
\end{figure}

\textbf{Price:}
Much previous literature has been triggered by the substantial fluctuations in crypto prices.
In this study we investigate 18 crypto prices in USD on daily time series.
Among these 18 cryptos, Bitcoin has been dominant and Reddcoin has the lowest price on balance as seen in Figure~\ref{fig:price}.
We characterise these price time series in Table~\ref{tab:price_char}.
Most of these coins (i.e., Bitcoin, Ethereum, Bitcoin Cash) have high fluctuations in price; while some coins (i.e., XRP, Blackcoin) tends to be steady.

\begin{figure}[!htb]
    \centering
    \includegraphics[width=1\textwidth]{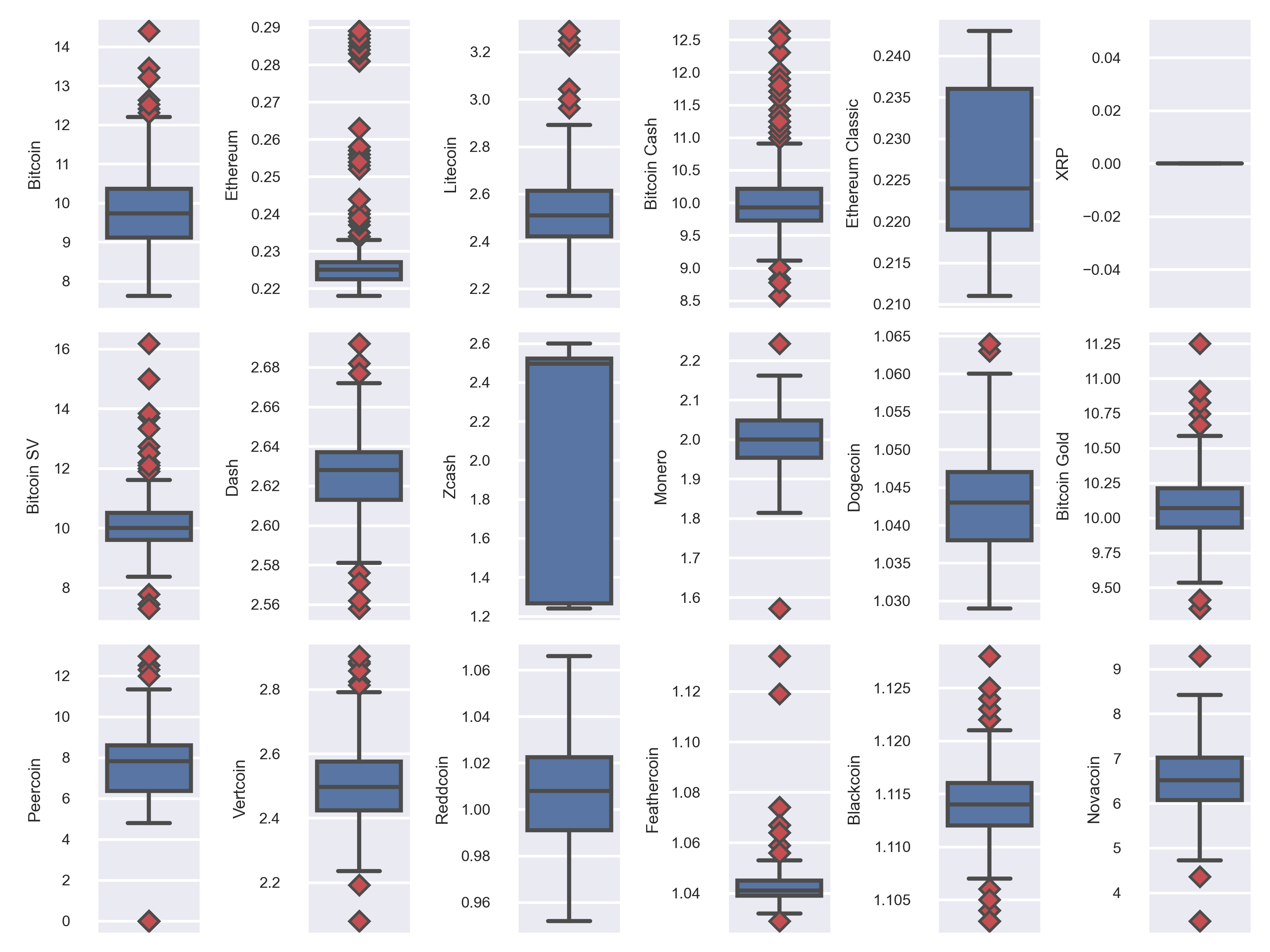}
    \caption{Actual block time in minutes.
    \protect \includegraphics[height=0.5cm]{images/qletlogo_tr.png} {\color{blue}\href{https://github.com/QuantLet/Blockchain\_mechanism/tree/main/Blockchain\_mechanism_plotting}{Blockchain\_mechanism\_plotting}}}
    \label{fig:block_time}
\end{figure}

\textbf{Actual block time:}
It is the mean time required in minutes for each day to create the next block.
In other words, it is the average amount of time for the day a user has to wait, after broadcasting their transaction, to see this transaction appear on the blockchain.
Some literature also refers it as confirmation time.
It can be considered as a service level indicator for cryptos which should be maintained by underlying mechanisms. 
Most of the coins discussed in this paper tend to have lower block time compared with Bitcoin as seen in Figure~\ref{fig:block_time}.
Also, many coins show outliers in observations and this can indicate that the extreme events appear in the blockchain system.
The underlying mechanisms can be ineffective to accommodate the current system demand.  
The distributional characteristics for time series of actual block time are presented in Table~\ref{tab:time_char}.
The data for XRP are missing but its designed block time is around 5 second per transaction.

\begin{figure}[!htb]
    \centering
    \includegraphics[width=1\textwidth]{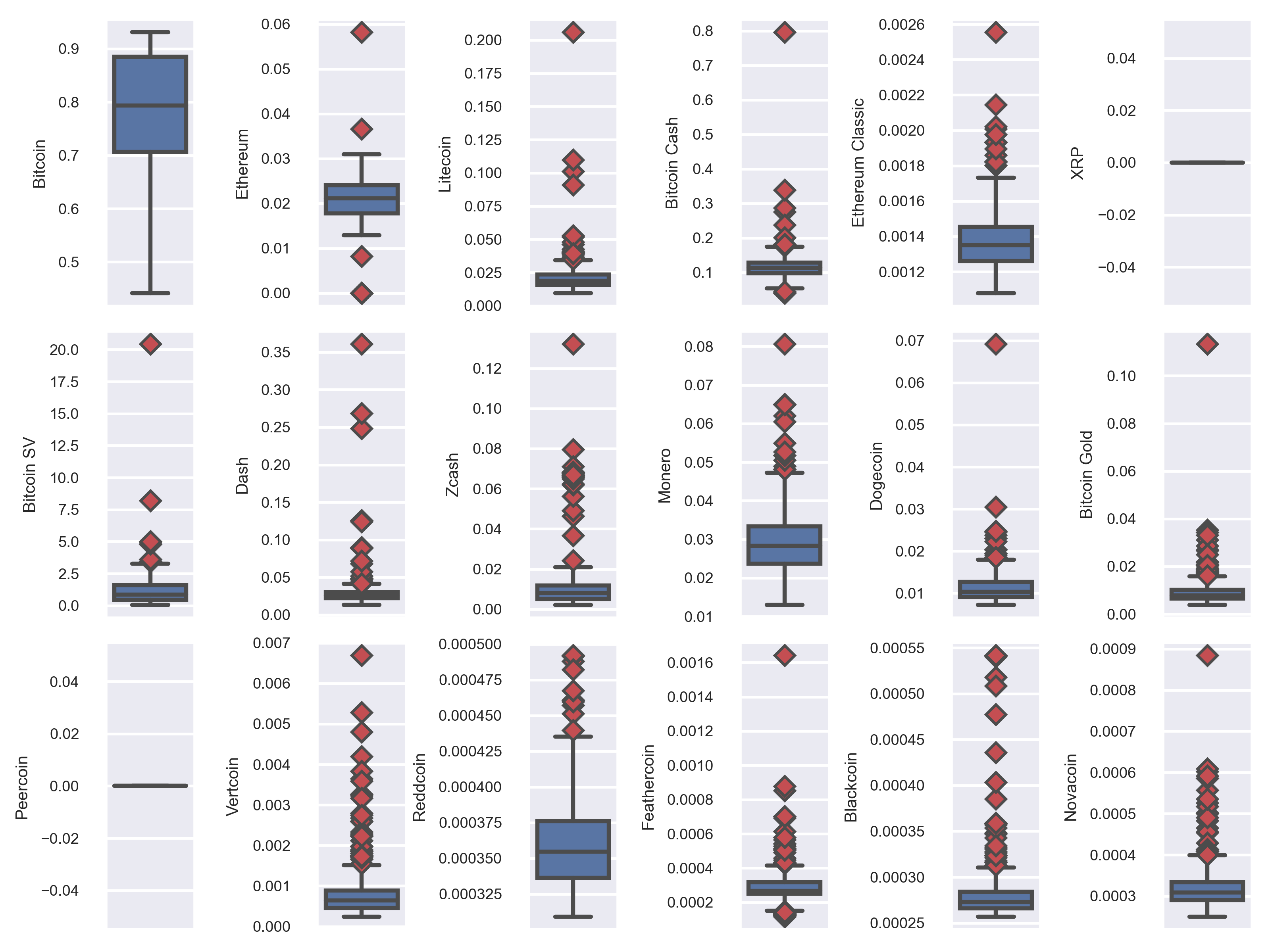}
    \caption{Actual block size in megabytes.
    \protect \includegraphics[height=0.5cm]{images/qletlogo_tr.png} {\color{blue}\href{https://github.com/QuantLet/Blockchain\_mechanism/tree/main/Blockchain\_mechanism_plotting}{Blockchain\_mechanism\_plotting}}}
    \label{fig:block_size}
\end{figure}

\textbf{Actual block size:}
It is defined as the average actual size "usage" of a single block in data storage for one day.
Since a block is is comprised of transaction data, it can represent the status of how a cryptocurrency mechanism allocates transactions to a block. 
In this study, as introduced in \hyperref[secIntro]{Section 1}, we consider it as an indicator for the stableness of scalability of a crypto.
In Figure~\ref{fig:block_size} shows that  most of the cryptos under study have smaller block size usage than Bitcoin, except Bitcoin SV. The plot also depicts that almost all the coins have outliers. 
These outliers can lead to the imbalance in transaction fee and reward which can influence the ecosystem of a crypto.
The characteristics for block size time series are shown in Table~\ref{tab:size_char}. 
XRP does not have typical blockchain structure, hence, there is no block size data in the study.
The data for Peercoin are missing.

\label{secMeth}
\section{Methodology}

In order to investigate the relationship between underlying blockchain mechanism of cryptocurrencies and the distributional characteristics of cryptos as a proxy of behaviour, we aim to group them into number of clusters and scrutinise the compositions of features in each group. These blockchain-based features manifest the underlying mechanism of how the cryptos operate transactions on their chains, and subsequently govern the price, actual block size and block time. As described in the previous section, we use the time series data of 18 different cryptos with a range of different mechanisms.

The time series data available for the cryptos is subject to numerous limitations. The most important one of them is that different coins were introduced at different time points, therefore, the data available for each coin has different lengths. For the clustering problems \citep{aghabozorgi2015time}, defining the distance metric between points in time series with various lengths is not conventional. For many analytical problems, this issue is easily tackled by truncating the time series to the shared sample period. We refrain from doing so because, in the analysis of cryptocurrency prices, the evolution of the data in time is highly crucial for an investigation in the short term and long term dynamics and therefore, truncating the time series would lead to loss of important information. Hence, we deal with the time series data of cryptos with different lengths and do not directly impose a distance metric on the input data points. 

Furthermore, characterising the behaviour of a time series in terms of a single quantitative attribute (such as range based volatility) has its own limitations. The chosen attribute usually captures the dynamics of time series in one particular aspect, which may not be sufficient to encompass an entire behaviour or introduces a biased assessment. This becomes particularly true in the problems of crypto classification and clustering where these attributes, used as a similarity measure, are very diverse, resulting in weak robustness in the results.  

To cope with these limitations, we resort to the characteristic based clustering method proposed by \cite{Hyndman2005}. It was recently applied by \cite{pele2020} for classifying cryptos in order to distinguish them from traditional assets. This methods recommends to incorporate various global measures describing the structural characteristics of a time series for a clustering problem. These global measures are obtained by applying statistical operations that best represent the underlying characteristics. Also, by extracting a set of measures from the original time series we simply bypass the issue of defining a distance metric. It's understood that the global measures are domain-specific. Employing a greedy search algorithm, \cite{Hyndman2005} selects the pivotal features in the clustering tasks. In our case, we import the experts' discretion on the choice of features as distributional characteristics which best represent the dynamics of cryptocurrencies.

We choose a variety of measures for our analysis. Starting from the first four moments and quantiles that characterises the distribution and symmetry of the data, we include the statistics for concluding the global structure such as global optimum, as well as the measures for long term dependencies, risk and noise. The selected features are mean, standard deviation, skewness, kurtosis, maximum, minimum, first quartile, median, third quartile, 1\% and 5\% extreme quantiles as a measure of downside risk, linear trend, intercept, autocorrelation for long term dependency, self-similarity using Hurst exponent and chaos using Lyaponav’s exponents.

We further extend the methodology by including the power spectrum of time series as an additional measure. The power spectrum is obtained in this work using Fast Fourier Transform (FFT). For computational ease, discrete fourier transform (DFT) has been formalised as a linear operator that maps the data points in a discrete input signal $X  \left\{x_{1}, x_{2}, \cdots, x_{n}\right\} $ to the frequency domain $f = \left\{f_{1}, f_{2}, \cdots f_{n}\right\}$.

For a given time series $X$ of $n$ time points, sine and cosine functions are used to get the coefficients $\omega_{n}=e^{-2 \pi i} / \eta$ and the frequencies are calculated using the matrix multiplication:

\begin{equation}
\left[\begin{array}{c}
f_{1} \\
f_{2} \\
f_{3} \\
\vdots \\
f_{n}
\end{array}\right]=\left[\begin{array}{ccccc}
1 & 1 & 1 & \cdots & 1 \\
1 & \omega_{n} & \omega_{n}^{2} & \cdots & \omega_{n}^{n-1} \\
1 & \omega_{n}^{2} & \omega_{n}^{4} & \cdots & \omega_{n}^{2(n-1)} \\
\vdots & \vdots & \vdots & \ddots & \vdots \\
1 & \omega_{n}^{n-1} & \omega_{n}^{2(n-1)} & \cdots & \omega_{n}^{(n-1)^{2}}
\end{array}\right]\left[\begin{array}{c}
x_{1} \\
x_{2} \\
x_{3} \\
\vdots \\
x_{n}
\end{array}\right]
\end{equation}

This matrix multiplication involves $\mathcal{O}(n^2)$ and makes DFT computationally expensive. FFT is a fast algorithm to compute DFT using only $\mathcal{O}(n\log n)$ operations \citep{brunton2019}. A simple \textit{fft} command in python computes the FFT of the given time signal. The power spectrum of this signal is the normalised squared magnitude of the \textit{f} and it indicates how much variance of the initial space each frequency explains \citep{brunton2019}. Including the power spectrum as a feature for characteristic based clustering allows capturing the variability in the time signal that is not explained by any other measure. 

Accumulating all the aforementioned features in a vector gives in a reduced dimensional representation of time series of each crypto. These vectors are then used to cluster the cryptos into groups using spectral clustering. Spectral clustering exploits the eigenvalues of similarity matrix to cluster and results in more balanced clusters than other techniques that were employed during the process. For details related to spectral clustering, the readers are recommended to follow the tutorial on spectral clustering by \cite{Luxburg2006}. The results of the above methodology are discussed in detail in the next section.

%
%
%
%
%
%
%
\label{sec_Res}
\section{Empirical Evidence}

In this section, we showcase the result from the characteristic based clustering individually on the crypto price and operational features--which are constructed with price, block size "scalability" and block time "service level" time series.
We explore the clustering results and classify them with the underlying mechanisms of the investigated 18 cryptos.
The 18 cryptos are: Bitcoin, Bitcoin Cash, Bitcoin Gold, Bitcoin SV, Blackcoin , Dash, Dogecoin, Ethereum, Ethereum Classic, Feathercoin, Litecoin, Monero, Novacoin, Peercoin, Reddcoin, Vertcoin, XRP, and Zcash.

We calculate the characteristics for each of these cryptos for prices, block size and block time separately. 
The results of all other attributes except the FFT are summarised in Tables \ref{tab:price_char}, \ref{tab:time_char}, \ref{tab:size_char} correspondingly in \hyperref[secApp]{Appendix}. 
Note that the data for XRP are not available for the block size and block time, and for Peercoin block size is missing as described before in \hyperref[secData]{Section 3}. 

After calculating the attributes and FFT power spectrum described in section \ref{secMeth}, the feature space is 216 dimensional (200 dimensional vector of power spectrum and 16 characteristics), visualisation of which is not possible. 
We project the feature space into a three dimensional space using principle component analysis (PCA), and the results of which are exhibited for an intuitive understanding. 
We discuss each of the clustering in detail below.
Moreover, in order to avoid a monopoly outcome and sustain a certain level of interpretability, we impose the maximum number of the clusters to avoid a single coin case in each cluster. 

\subsection{Clustering with crypto prices}
Table \ref{tab:price_char} shows that as expected, Bitcoin has the highest average price and highest standard deviation, due to high magnitude of its prices.
The VaR99 and VaR95 for Bitcoin are, however, very low, showing a low downside risk of Bitcoin.
On the contrary, Bitcoin Cash, Bitcoin SV, Bitcoin Gold and Zcash all show high value at risk. 
This could be due to low persistence of risk shocks \citep{deSouza2019, KATSIAMPA201935}.
The high positive coefficients of self similarity for all the coins show high dependency on the previous time values. The high autocorrelation further confirms the presence of long term dependencies of the time series. The Lyaponov exponent as a measure of chaos is greater than 0 for all the time series which shows unstable dynamics throughout the prices of cryptos. 

The characteristics of Dogecoin in Table \ref{tab:price_char} assume very low values, unlike any other coin, because the prices of Dogecoin are very low, despite it being a popular coin. This can be due to high supply of the coin with no limit on the total number of coins created. 
The coin also has no technical innovations, which is considered as one of the reasons why the coin has such small price. 
Hence, the uncontrolled underlying mechanism of the coin has significant impact on the prices, despite the high trading volumes of the coin. 
Same can be concluded for XRP and Reddcoin, which also have a very high maximum supply that is reflected in their very low prices. 

Using characteristic based clustering on price time series, we have the result with 5 clusters as below:
\begin{enumerate}
\setcounter{enumi}{-1}
    \item  Bitcoin, Dash
    \item  Bitcoin SV, Zcash
    \item  Bitcoin Cash, Bitcoin Gold
    \item  Ethereum, Litecoin, XRP, Monero, Peercoin, Vertcoin, Reddcoin, Feathercoin, Blackcoin
    \item Ethereum Classic, Dogecoin, Novacoin
\end{enumerate}

\begin{figure}[!htb]
    \centering
    \includegraphics[width=0.8\textwidth]{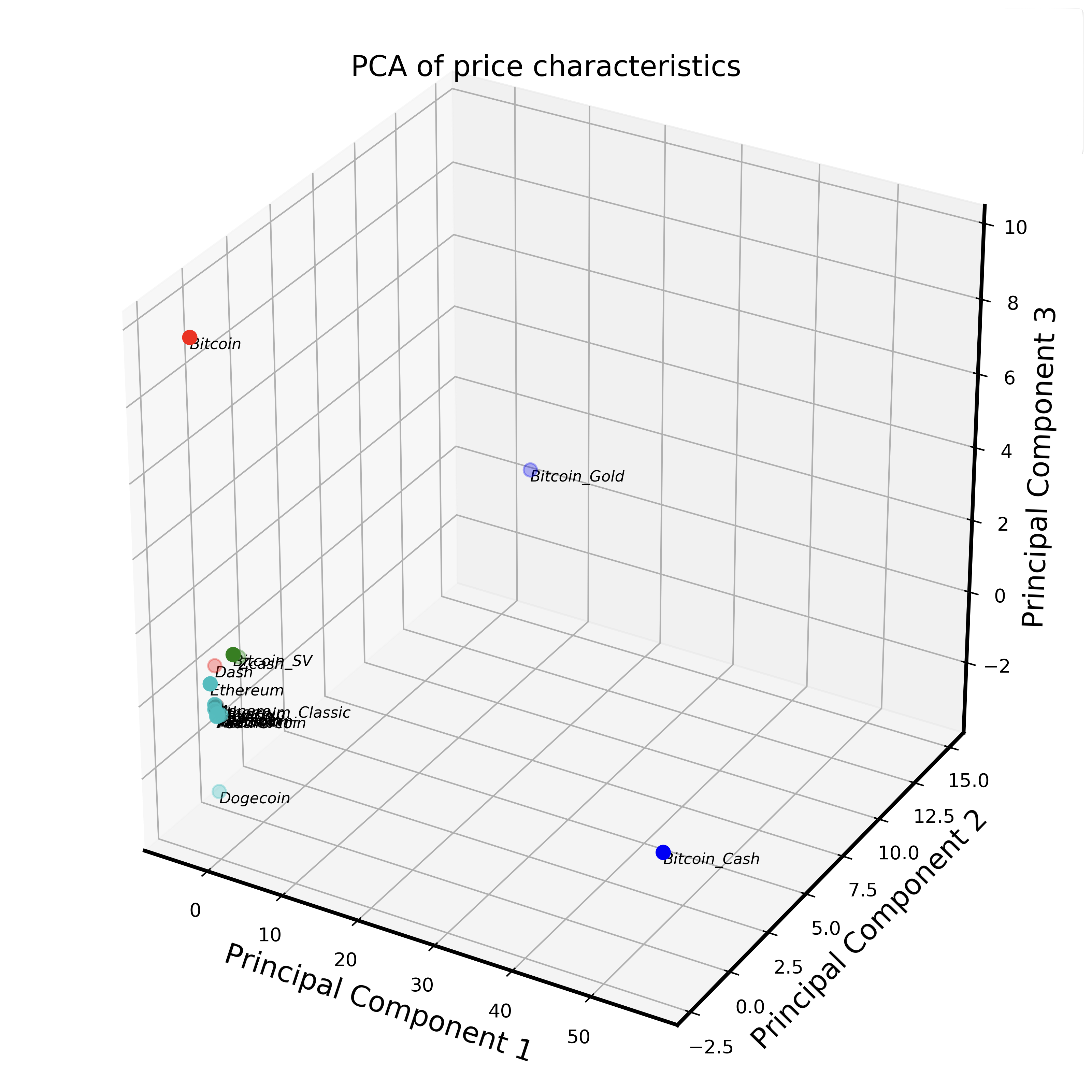}
    \caption{Visualisation of five clusters \textbf{{\color{red}0}, {\color{green}1}, {\color{blue}2}, {\color{black}3}, {\color{cyan}4}} of cryptos based on the prices
    \protect \includegraphics[height=0.5cm]{images/qletlogo_tr.png} {\color{blue}\href{https://github.com/QuantLet/Blockchain\_mechanism/tree/main/Blockchain\_mechanism\_clustering}{Blockchain\_mechanism\_clustering}}
    }
    \label{fig:price_PCA}
\end{figure}

Most of coins are close to each others in a three-dimensional space, as seen in Figure~\ref{fig:price_PCA}.
Except Dash, all the altcoins are in a different clusters than Bitcoin.
Bitcoin Cash and Bitcoin Gold, which principally inherit the protocol architecture from Bitcoin, are clustered together, but not centred around with other coins.
However, Bitcoin SV--which is a fork from Bitcoin Cash and mainly increases the designed block size to lower the transaction fee as a main software change--is not in the same cluster.
This indicates that even as a crypto adopts a similar blockchain mechanism with the other crypto, it might have different price dynamics than its origin.

XRP, Monero, Peercoin, Reddcoin, and Blackcoin which apply significantly different blockchain protocols in their governance types and consensus mechanisms are in the same cluster.
Specifically, XRP, Monero and Peercoin are private based blockchain which possesses a stronger moderator to control the entrants (users or investors) to their network.
Peercoin, Reddcoin, and Blackcoin, instead of using PoW as their consensus mechanisms, employ PoS which does not depends on miners' effort to create a block.
So that, coin supply and demand can reach an equilibrium without the interference of miners, which leads to higher transaction costs.
Moreover, the forks from Litecoin--Vertcoin, Reddcoin and Feathercoin are within the same cluster with Litecoin.

Ethereum Classic is, in fact, the version of Ethereum that existed before the hard fork of Ethereum resulting after the DAO attack, but it is not within the cluster with Ethereum.

\subsection{Clustering with actual block time}
The block time here is measured in minutes. Likewise, we apply the characteristic based clustering on the data and conclude them into 5 clusters as below.
\begin{enumerate}
\setcounter{enumi}{-1}
    \item  Dogecoin, Feathercoin
    \item  Ethereum, Litecoin, Ethereum Classic, Dash, Zcash, Monero, Blackcoin
    \item  Bitcoin, Bitcoin Cash, Vertcoin
    \item  Bitcoin SV, Bitcoin Gold, Novacoin
    \item  Peercoin, Reddcoin
\end{enumerate}

The result is correspondingly visualised in Figure~\ref{fig:block_time}. 
The figure shows that Peercoin and Reddcoin lie far away from other coins (marked by cyan cluster). 
They are clustered in the same group because they both use PoS and their initial block takes the maximum time to be added, as shown by the maximum and intercept characteristics in Table~\ref{tab:time_char}. 
This shows that even though the coins have lower actual block time later (with low mean), their behaviour is still the similar, resulting them in the same cluster.
Also, the cryptos using PoS tend to lower the complexity of their hashing algorithms since it is not required for miners to spend computational effort on them. 
The difficulty adjustment algorithms of theirs are purely used as a mechanism for maintaining the certain service level for users without considering hashrate from miners.
Their block time performance is relatively stable after the initialisation.
Here we emphasise that the initial price, block time and block size that are usually characterised by the underlying mechanism play a pivotal role in determining the price behaviour of cryptos. 
This is why we did not truncate the time series, as mentioned in the \hyperref[secMeth]{Section 4}.

\begin{figure}[!htb]
    \centering
    \includegraphics[width=0.8\textwidth]{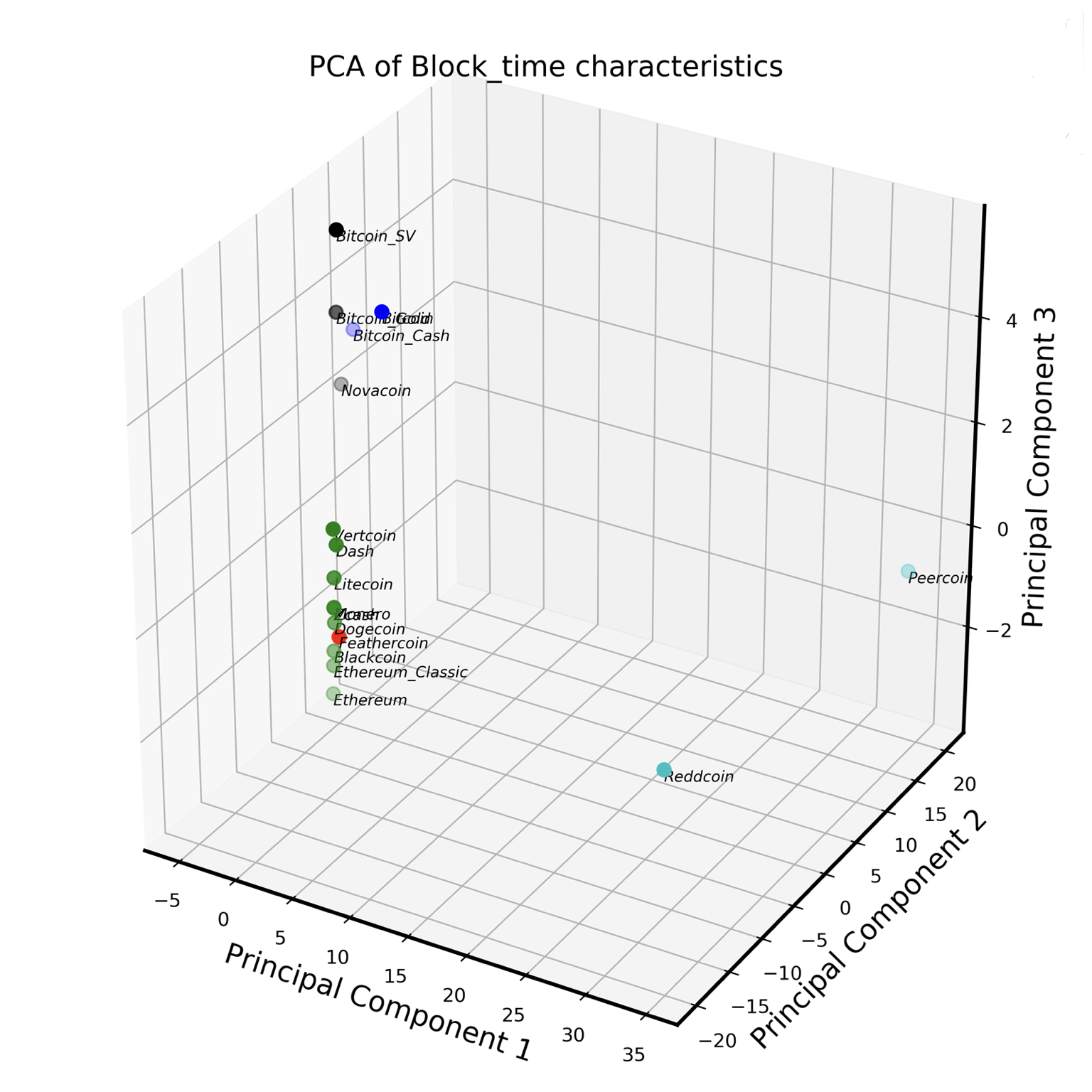}
    \caption{Visualisation of five clusters \textbf{{\color{red}0}, {\color{green}1}, {\color{blue}2}, {\color{black}3}, {\color{cyan}4}} of cryptos based on block time
    \protect \includegraphics[height=0.5cm]{images/qletlogo_tr.png} {\color{blue}\href{https://github.com/QuantLet/Blockchain\_mechanism/tree/main/Blockchain\_mechanism\_clustering}{Blockchain\_mechanism\_clustering}}}
    \label{fig:blocktime_PCA}
\end{figure}

Though Bitcoin, Bitcoin Gold, Bitcoin Cash and Bitcoin SV are not completely grouped into the same cluster, they are close to each others in the three dimensional space as seen in Figure~\ref{fig:block_time}. 
They apply the same hashing algorithm--SHA-256 and also with the same expected block time for their difficulty adjustment algorithms. 
Let's call attention to forks again. 
Dogecoin and Feathercoin are both forked from Litecoin with the Script-based hashing algorithm and difficulty adjustment frequency after large number of blocks--240 and 504 blocks. 
Litecoin is in a different cluster because the frequency is much higher as 2016 blocks.
Given the cryptos forked from the same origin coins, their block time can be found in the same group, likewise Ethereum and Ethereum Classic. 

\subsection{Clustering with actual block size}
As previously done for price and block time, we use the characteristics based clustering and grouped these cryptos into 5 clusters according to the characteristics of their time series.
The block size here is measured in bytes for a better data representation.
As stated before in \hyperref[secData]{Section 3}, XRP and Peercoin data are missing due to the mechanism design and incomplete data from the source, respectively.
The clustering result is shown as below and the corresponding visualisation is in Figure~\ref{fig:block_size}.
\begin{enumerate}
\setcounter{enumi}{-1}
    \item  Zcash, Bitcoin Gold, Reddcoin, Novacoin
    \item  Ethereum, Ethereum Classic, Dogecoin
    \item  Bitcoin Cash, Bitcoin SV
    \item  Bitcoin, Dash, Monero, Feathercoin
    \item  Litecoin, Vertcoin, Blackcoin
\end{enumerate}

\begin{figure}[!htb]
    \centering
    \includegraphics[width=0.8\textwidth]{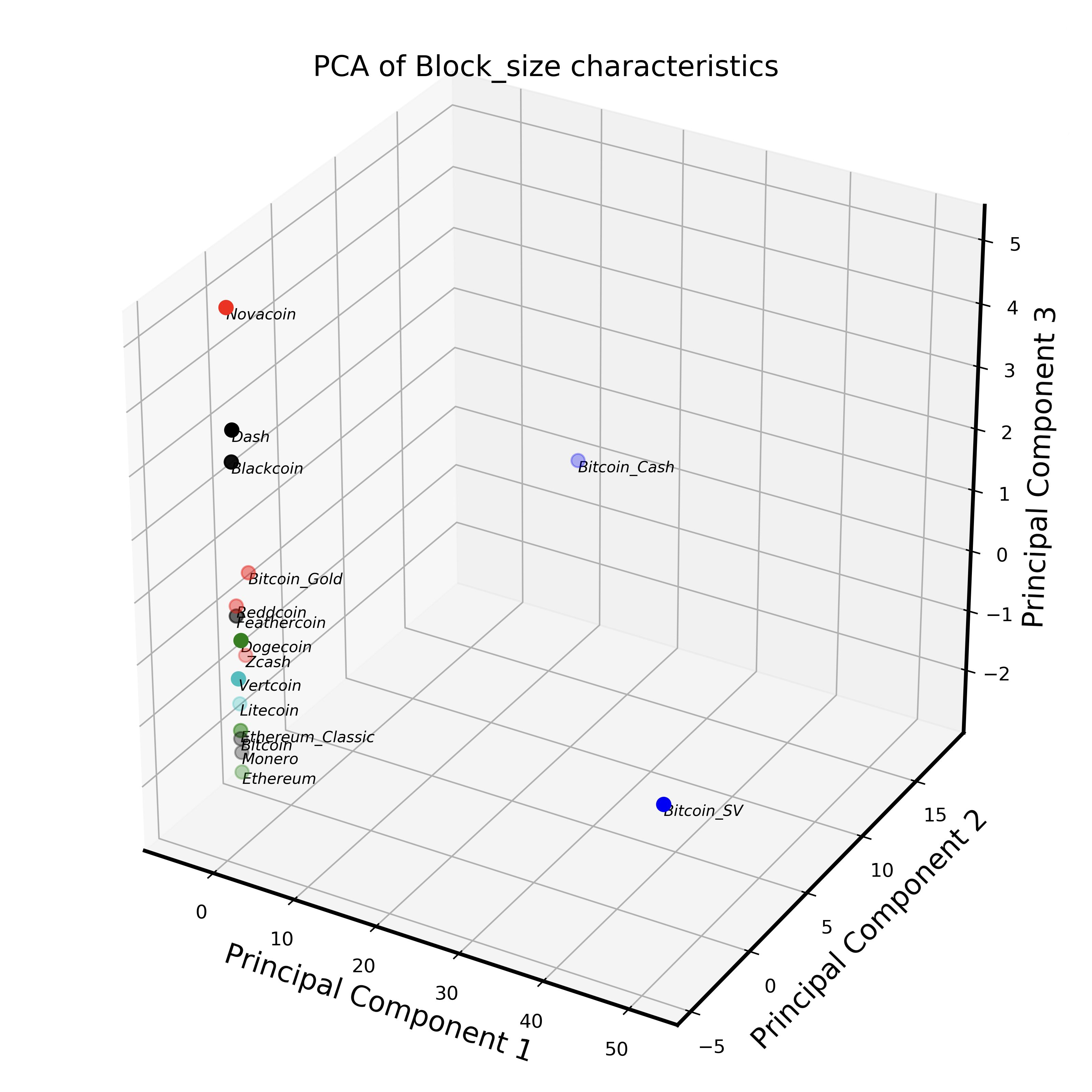}
    \caption{Visualisation of five clusters \textbf{{\color{red}0}, {\color{green}1}, {\color{blue}2}, {\color{black}3}, {\color{cyan}4}} of cryptos based on block size
    \protect \includegraphics[height=0.5cm]{images/qletlogo_tr.png} {\color{blue}\href{https://github.com/QuantLet/Blockchain\_mechanism/tree/main/Blockchain\_mechanism\_clustering}{Blockchain\_mechanism\_clustering}}}
    \label{fig:Blocksize_PCA}
\end{figure}

The actual block size (usage) of these cryptos does rarely meet their designed block size limit (capacity), except for Bitcoin that it nearly outstretches its limit, 1 megabyte, as seen in Table~\ref{tab:size_char}.
In this case, it raises an issue: Can increasing crypto's block size limit  improves scalability?
For example, Bitcoin SV enlarges dramatically its limit to 128 megabytes but it is out of the necessity for such a design.
Likewise, Bitcoin Cash, which Bitcoin SV forks from, has its limit as 32 megabytes.
These two coins are, therefore, clustered together.
Moreover, instead of having a static block size limit, Ethereum and Ethereum Classic grouped in the same cluster apply block gas limit, which is the energy consumption limit for a block, to adaptively regulate its block size.
Both Monero and Blackcoin have a dynamic mechanisms to control the block size, however, it does not represent in the clustering result.

\label{secCol}
\section{Conclusion}
In this paper we investigate the relationship between crypto behaviours and their underlying mechanisms.
We specify the crypto behaviour with their price and operational features defined by actual block time and block size. We calculate the distributional characteristics to define the behaviour of time series. Using a characteristics based spectral clustering technique, we cluster the selected coins into a number of clusters and scrutinise the blockchain mechanism in each group. We find that the underlying mechanism of cryptos are reflected in the clustering results. We observe that cryptos forked from same origin and same consensus mechanism tend to become part of same clustering group. Furthermore, the clusters obtained by the time series of block time have same hashing algorithms and difficulty adjustment algorithms. Also, a similar nature (static or dynamic) of block size was observed within clusters obtained by the time series of actual block size. We conclude with empirical evidence that the crypto behaviour is indeed linked with their blockchain protocol architectures. As a result, cryptocurrency users and investors can have a better understanding and explanation of price and operational features through cryptocurrency mechanism. In the future research, we would elaborate the relation of price and operational features to underlying mechanism with an economic model and conduct relevant simulations. We would also like to investigate the impact of versions revisions on the dynamics of cryptos.

\bibliography{references.bib}

\begin{thebibliography}{20}
\providecommand{\natexlab}[1]{#1}
\providecommand{\url}[1]{\texttt{#1}}
\expandafter\ifx\csname urlstyle\endcsname\relax
  \providecommand{\doi}[1]{doi: #1}\else
  \providecommand{\doi}{doi: \begingroup \urlstyle{rm}\Url}\fi

\bibitem[Aghabozorgi et~al.(2015)Aghabozorgi, Shirkhorshidi, and
  Wah]{aghabozorgi2015time}
S.~Aghabozorgi, A.~S. Shirkhorshidi, and T.~Y. Wah.
\newblock Time-series clustering--a decade review.
\newblock \emph{Information Systems}, 53:\penalty0 16--38, 2015.

\bibitem[Blau et~al.(2020)Blau, Griffith, and Whitby]{Blau2020}
B.~Blau, T.~Griffith, and R.~Whitby.
\newblock {Comovement in the Cryptocurrency Market}.
\newblock \emph{Economics Bulletin}, 40\penalty0 (1):\penalty0 448--455, 2020.

\bibitem[Brunton and Kutz(2019)]{brunton2019}
S.~L. Brunton and J.~N. Kutz.
\newblock \emph{Data-Driven Science and Engineering: Machine Learning,
  Dynamical Systems, and Control}.
\newblock Cambridge University Press, 2019.
\newblock \doi{10.1017/9781108380690}.

\bibitem[Buterin(2014)]{buterin2014next}
V.~Buterin.
\newblock A next-generation smart contract and decentralized application
  platform.
\newblock \emph{white paper}, 2014.

\bibitem[Ciaian et~al.(2016)Ciaian, Rajcaniova, and d’Artis Kancs]{Pavel2016}
P.~Ciaian, M.~Rajcaniova, and d’Artis Kancs.
\newblock The economics of bitcoin price formation.
\newblock \emph{Applied Economics}, 48\penalty0 (19):\penalty0 1799--1815,
  2016.
\newblock \doi{10.1080/00036846.2015.1109038}.

\bibitem[de~Souza(2019)]{deSouza2019}
M.~de~Souza.
\newblock Var and persistence of risk shocks in cryptocurrencies market.
\newblock \emph{The Empirical Economics Letters}, 18:\penalty0 1--12, 04 2019.

\bibitem[Garriga et~al.(2020)Garriga, Dalla~Palma, Arias, Derenzis, Pareschi,
  and Tamburri]{Garriga2020}
M.~Garriga, S.~Dalla~Palma, M.~Arias, A.~Derenzis, R.~Pareschi, and
  D.~Tamburri.
\newblock Blockchain and cryptocurrencies: A classification and comparison of
  architecture drivers.
\newblock \emph{Concurrency and Computation Practice and Experience}, 10 2020.
\newblock \doi{10.1002/cpe.5992}.

\bibitem[Guo et~al.(2018)Guo, Tao, and Härdle]{Li2018}
L.~Guo, Y.~Tao, and W.~Härdle.
\newblock Understanding latent group structure of cryptocurrencies market: A
  dynamic network perspective.
\newblock \emph{SSRN Electronic Journal}, 01 2018.
\newblock \doi{10.2139/ssrn.3658206}.

\bibitem[Guo and Donev(2020)]{guo2020bibliometrics}
X.~Guo and P.~Donev.
\newblock Bibliometrics and network analysis of cryptocurrency research.
\newblock \emph{Journal of Systems Science and Complexity}, pages 1--26, 2020.

\bibitem[Hou et~al.(2020)Hou, Wang, Chen, and Härdle]{Hou2020pricing}
A.~J. Hou, W.~Wang, C.~Y.~H. Chen, and W.~K. Härdle.
\newblock {Pricing Cryptocurrency Options*}.
\newblock \emph{Journal of Financial Econometrics}, 18\penalty0 (2):\penalty0
  250--279, 05 2020.
\newblock ISSN 1479-8409.
\newblock \doi{10.1093/jjfinec/nbaa006}.
\newblock URL \url{https://doi.org/10.1093/jjfinec/nbaa006}.

\bibitem[Härdle et~al.(2020)Härdle, Harvey, and
  Reule]{haerdle2020understanding}
W.~K. Härdle, C.~R. Harvey, and R.~C.~G. Reule.
\newblock {Understanding Cryptocurrencies*}.
\newblock \emph{Journal of Financial Econometrics}, 18\penalty0 (2):\penalty0
  181--208, 02 2020.
\newblock ISSN 1479-8409.
\newblock \doi{10.1093/jjfinec/nbz033}.
\newblock URL \url{https://doi.org/10.1093/jjfinec/nbz033}.

\bibitem[Iwamura et~al.(2019)Iwamura, Kitamura, Matsumoto, and
  Saito]{Iwamura2019}
M.~Iwamura, Y.~Kitamura, T.~Matsumoto, and K.~Saito.
\newblock Can we stabilize the price of a cryptocurrency?: Understanding the
  design of bitcoin and its potential to compete with central bank money.
\newblock \emph{Hitotsubashi Journal of Economics}, 60\penalty0 (1):\penalty0
  41--60, 2019.

\bibitem[Katsiampa et~al.(2019)Katsiampa, Corbet, and Lucey]{KATSIAMPA201935}
P.~Katsiampa, S.~Corbet, and B.~Lucey.
\newblock High frequency volatility co-movements in cryptocurrency markets.
\newblock \emph{Journal of International Financial Markets, Institutions and
  Money}, 62:\penalty0 35 -- 52, 2019.
\newblock ISSN 1042-4431.
\newblock \doi{https://doi.org/10.1016/j.intfin.2019.05.003}.
\newblock URL
  \url{http://www.sciencedirect.com/science/article/pii/S104244311930023X}.

\bibitem[King and Nadal(2012)]{king2012ppcoin}
S.~King and S.~Nadal.
\newblock Ppcoin: Peer-to-peer crypto-currency with proof-of-stake.
\newblock \emph{self-published paper, August}, 19:\penalty0 1, 2012.

\bibitem[Pele et~al.(2020)Pele, Wesselh{\"o}fft, H{\"a}rdle, Kolossiatis, and
  Yatracos]{pele2020}
D.~T. Pele, N.~Wesselh{\"o}fft, W.~K. H{\"a}rdle, M.~Kolossiatis, and
  Y.~Yatracos.
\newblock A statistical classification of cryptocurrencies.
\newblock \emph{Available at SSRN 3548462}, 2020.

\bibitem[Sovbetov(2018)]{sovbetov2018factors}
Y.~Sovbetov.
\newblock Factors influencing cryptocurrency prices: Evidence from bitcoin,
  ethereum, dash, litcoin, and monero.
\newblock \emph{Journal of Economics and Financial Analysis}, 2\penalty0
  (2):\penalty0 1--27, 2018.

\bibitem[Trimborn and Härdle(2018)]{TRIMBORN2018107}
S.~Trimborn and W.~K. Härdle.
\newblock Crix an index for cryptocurrencies.
\newblock \emph{Journal of Empirical Finance}, 49:\penalty0 107 -- 122, 2018.
\newblock ISSN 0927-5398.
\newblock \doi{https://doi.org/10.1016/j.jempfin.2018.08.004}.
\newblock URL
  \url{http://www.sciencedirect.com/science/article/pii/S0927539818300616}.

\bibitem[von Luxburg(2006)]{Luxburg2006}
U.~von Luxburg.
\newblock A tutorial on spectral clustering.
\newblock Technical Report 149, Max Planck Institute for Biological
  Cybernetics, Tübingen, Aug. 2006.

\bibitem[Wang et~al.(2005)Wang, Smith-Miles, and Hyndman]{Hyndman2005}
X.~Wang, K.~Smith-Miles, and R.~Hyndman.
\newblock Characteristic-based clustering for time series data.
\newblock \emph{Data Mining and Knowledge Discovery}, 13:\penalty0 335--364,
  2005.

\bibitem[Zimmerman(2020)]{zimmerman2020blockchain}
P.~Zimmerman.
\newblock Blockchain structure and cryptocurrency prices.
\newblock \emph{Bank of England Working Paper}, 2020.

\end{thebibliography}
\label{secApp}
\section*{Appendix}

\begin{table}[H]
\footnotesize
\singlespacing
\caption{Characteristics of prices of different cryptocurrencies}
\label{tab:price_char}
\resizebox{\textwidth}{!}{%
\begin{tabular}{rrrrrrr}
\hline
\textbf{Characteristic} & \textbf{Bitcoin} & \textbf{Ethereum} & \textbf{Litecoin} & \textbf{\begin{tabular}[c]{@{}l@{}}Bitcoin\\ Cash\end{tabular}} & \textbf{\begin{tabular}[c]{@{}l@{}}Ethereum\\ Classic\end{tabular}} & \textbf{XRP} \\ \hline
\textbf{mean} & 2659.127 & 178.966 & 34.394 & 537.723 & 9.381 & 0.192 \\
\textbf{standard\_deviation} & 3798.466 & 222.452 & 48.645 & 509.244 & 7.827 & 0.302 \\
\textbf{skewness} & 1.338 & 1.950 & 2.389 & 2.322 & 1.491 & 4.193 \\
\textbf{kurtosis} & 0.672 & 4.654 & 7.272 & 6.157 & 2.239 & 29.471 \\
\textbf{maximum} & 19401.000 & 1356.000 & 352.799 & 3526.000 & 43.765 & 3.649 \\
\textbf{minimum} & 0.050 & 0.401 & 0.032 & 58.626 & 0.687 & 0.003 \\
\textbf{lowerquant} & 20.193 & 7.975 & 3.153 & 233.404 & 4.364 & 0.007 \\
\textbf{median} & 455.892 & 136.557 & 8.618 & 324.646 & 6.571 & 0.024 \\
\textbf{upperquant} & 5128.000 & 250.965 & 53.128 & 620.947 & 13.813 & 0.291 \\
\textbf{VaR99} & 0.062 & 0.578 & 0.040 & 107.426 & 0.809 & 0.004 \\
\textbf{VaR95} & 0.393 & 0.696 & 0.072 & 129.491 & 1.105 & 0.005 \\
\textbf{slope} & 2.781 & 0.163 & 0.032 & -0.876 & -0.002 & 0.000 \\
\textbf{intercept} & 0.050&	2.820&	0.033&	63.765&	0.892 & 0.006\\
\textbf{autocorrelation} & 0.998 & 0.998 & 0.997 & 0.992 & 0.994 & 0.991 \\
\textbf{self\_similarity} & 1.574 & 1.611 & 1.596 & 1.609 & 1.564 & 1.551 \\
\textbf{chaos} & 0.088 & 0.093 & 0.091 & 0.086 & 0.087 & 0.085 \\ \hline
\end{tabular}%
}
\resizebox{\textwidth}{!}{%
\begin{tabular}{rrrrrrr}
\hline
\textbf{Characteristic} & \textbf{\begin{tabular}[c]{@{}l@{}}Bitcoin\\ SV\end{tabular}} & \textbf{Dash}    & \textbf{Zcash} & \textbf{Monero} & \textbf{Dogecoin} & \textbf{\begin{tabular}[c]{@{}l@{}}Bitcoin \\ Gold\end{tabular}} \\ \hline
\textbf{mean} & 145.401 & 113.910 & 135.596 & 57.588 & 0.006 & 43.167 \\
\textbf{standard\_deviation} & 66.784 & 187.915 & 125.654 & 75.569 & 0.193 & 70.420 \\
\textbf{skewness} & 0.678 & 3.126 & 1.756 & 2.145 & 49.692 & 2.879 \\
\textbf{kurtosis} & 0.079 & 11.777 & 3.208 & 5.300 & 2469.511 & 8.351 \\
\textbf{maximum} & 370.647 & 1436.000 & 728.159 & 439.391 & 9.608 & 513.293 \\
\textbf{minimum} & 52.683 & 0.516 & 23.940 & 0.233 & 0.000 & 5.093 \\
\textbf{lowerquant} & 87.323 & 3.950 & 50.251 & 1.100 & 0.000 & 9.710 \\
\textbf{median} & 135.217 & 66.508 & 72.251 & 44.090 & 0.001 & 15.869 \\
\textbf{upperquant} & 191.739 & 133.239 & 199.807 & 84.834 & 0.003 & 29.706 \\
\textbf{VaR99} & 53.377 & 0.711 & 27.767 & 0.272 & 0.000 & 5.357 \\
\textbf{VaR95} & 62.111 & 1.833 & 31.842 & 0.417 & 0.000 & 6.604 \\
\textbf{slope} & 0.218 & 0.083 & -0.134 & 0.053 & 0.000 & -0.147 \\
\textbf{intercept} & 111.700 &	1.380&	286.297	&1.911&	0.000&	513.293 \\
\textbf{autocorrelation} & 0.990 & 0.997 & 0.995 & 0.997 & 0.002 & 0.961 \\
\textbf{self\_similarity} & 1.628 & 1.642 & 1.573 & 1.577 & 1.024 & 1.431 \\
\textbf{chaos} & 0.077 & 0.090 & 0.092 & 0.091 & 0.086 & 0.073 \\ \hline
\end{tabular}%
}
\resizebox{\textwidth}{!}{%
\begin{tabular}{rrrrrrr}
\hline
\textbf{Characteristic} & \textbf{\begin{tabular}[c]{@{}l@{}}Peer   \\coin\end{tabular}} & \textbf{\begin{tabular}[c]{@{}l@{}}Vertcoin\\ \end{tabular}} & \textbf{\begin{tabular}[c]{@{}l@{}}Redd-\\ coin\end{tabular}} & \textbf{\begin{tabular}[c]{@{}l@{}}Feather-\\ coin\end{tabular}} & \textbf{\begin{tabular}[c]{@{}l@{}}Black-\\ coin\end{tabular}} & \textbf{\begin{tabular}[c]{@{}l@{}}Nova-\\ coin\end{tabular}} \\ \hline
\textbf{mean} & 1.004 & 0.670 & 0.001 & 0.062 & 0.095 & 2.185 \\
\textbf{standard\_deviation} & 1.238 & 1.319 & 0.003 & 0.102 & 0.127 & 2.989 \\
\textbf{skewness} & 2.511 & 3.637 & 4.175 & 3.379 & 3.397 & 3.102 \\
\textbf{kurtosis} & 7.017 & 14.792 & 24.526 & 17.172 & 15.251 & 12.916 \\
\textbf{maximum} & 9.118 & 9.386 & 0.029 & 1.203 & 1.108 & 24.777 \\
\textbf{minimum} & 0.110 & 0.006 & 0.000 & 0.002 & 0.014 & 0.078 \\
\textbf{lowerquant} & 0.291 & 0.043 & 0.000 & 0.008 & 0.030 & 0.507 \\
\textbf{median} & 0.445 & 0.237 & 0.001 & 0.019 & 0.045 & 0.901 \\
\textbf{upperquant} & 1.275 & 0.626 & 0.001 & 0.072 & 0.088 & 3.301 \\
\textbf{VaR99} & 0.125 & 0.009 & 0.000 & 0.003 & 0.015 & 0.156 \\
\textbf{VaR95} & 0.168 & 0.015 & 0.000 & 0.004 & 0.020 & 0.187 \\
\textbf{slope} & 0.000 & 0.000 & 0.000 & 0.000 & 0.000 & -0.001 \\
\textbf{intercept} & 0.382&	6.315&	0.000&	0.559&	0.035&	0.078\\
\textbf{autocorrelation} & 0.993 & 0.992 & 0.988 & 0.983 & 0.993 & 0.994 \\
\textbf{self\_similarity} & 1.577 & 1.603 & 1.548 & 1.523 & 1.537 & 1.596 \\
\textbf{chaos} & 0.088 & 0.085 & 0.079 & 0.078 & 0.084 & 0.091 \\ \hline
\end{tabular}%
}
\end{table}

\begin{table}[H]
\footnotesize
\singlespacing
\caption{Characteristics of Block time of different cryptocurrencies}
\label{tab:time_char}
\resizebox{\textwidth}{!}{%
\begin{tabular}{rrrrrrr}
\hline
\textbf{Characteristic} & \textbf{Bitcoin} & \textbf{Ethereum} & \textbf{Litecoin} & \textbf{\begin{tabular}[c]{@{}l@{}}Bitcoin\\ Cash\end{tabular}} & \textbf{\begin{tabular}[c]{@{}l@{}}Ethereum\\ Classic\end{tabular}} & \textbf{XRP} \\ \hline
\textbf{mean} & 10.453 & 0.257 & 2.507 & 11.167 & 0.246 & NA \\
\textbf{standard\_deviation} & 8.814 & 0.045 & 0.385 & 11.009 & 0.032 & NA \\
\textbf{skewness} & 21.779 & 3.098 & 5.003 & 11.597 & 5.144 & NA \\
\textbf{kurtosis} & 701.717 & 11.987 & 54.589 & 160.209 & 61.066 & NA \\
\textbf{maximum} & 360.000 & 0.509 & 8.521 & 205.714 & 0.800 & NA \\
\textbf{minimum} & 2.081 & 0.208 & 0.149 & 1.275 & 0.153 & NA \\
\textbf{lowerquant} & 8.623 & 0.235 & 2.357 & 9.664 & 0.235 & NA \\
\textbf{median} & 9.474 & 0.241 & 2.474 & 9.931 & 0.238 & NA \\
\textbf{upperquant} & 10.435 & 0.268 & 2.599 & 10.360 & 0.242 & NA \\
\textbf{VaR99} & 5.923 & 0.220 & 1.710 & 2.331 & 0.215 & NA \\
\textbf{VaR95} & 7.129 & 0.222 & 2.111 & 8.479 & 0.218 & NA \\
\textbf{slope} & -0.001 & 0.000 & 0.000 & -0.007 & 0.000 & NA \\
\textbf{intercept} & 102.857 & 0.208 & 0.149 & 160.000 & 0.208 & NA \\
\textbf{autocorrelation} & 0.494 & 0.981 & 0.705 & 0.395 & 0.818 & NA \\
\textbf{self\_similarity} & 1.027 & 1.522 & 0.787 & 0.704 & 1.249 & NA \\
\textbf{chaos} & 0.012 & 0.070 & 0.012 & 0.003 & 0.068 & NA \\ \hline
\end{tabular}%
}
\resizebox{\textwidth}{!}{%
\begin{tabular}{rrrrrrr}
\hline
\textbf{Characteristic} & \textbf{\begin{tabular}[c]{@{}l@{}}Bitcoin\\ SV\end{tabular}} & \textbf{Dash} & \textbf{Zcash} & \textbf{Monero} & \textbf{Dogecoin} & \textbf{\begin{tabular}[c]{@{}l@{}}Bitcoin \\ Gold\end{tabular}} \\ \hline
\textbf{mean} & 10.195 & 2.659 & 2.409 & 1.686 & 1.048 & 9.823 \\
\textbf{standard\_deviation} & 1.639 & 0.805 & 0.345 & 0.541 & 0.043 & 0.741 \\
\textbf{skewness} & 12.504 & 19.831 & -3.025 & 3.258 & -9.220 & -5.375 \\
\textbf{kurtosis} & 221.950 & 409.827 & 7.261 & 57.807 & 222.460 & 60.686 \\
\textbf{maximum} & 40.000 & 22.500 & 2.618 & 10.992 & 1.288 & 11.250 \\
\textbf{minimum} & 7.310 & 0.348 & 1.240 & 0.829 & 0.100 & 0.254 \\
\textbf{lowerquant} & 9.600 & 2.609 & 2.487 & 1.025 & 1.038 & 9.664 \\
\textbf{median} & 10.000 & 2.623 & 2.509 & 1.951 & 1.044 & 9.931 \\
\textbf{upperquant} & 10.511 & 2.637 & 2.531 & 2.020 & 1.050 & 10.141 \\
\textbf{VaR99} & 8.361 & 2.476 & 1.248 & 0.947 & 0.980 & 7.767 \\
\textbf{VaR95} & 9.034 & 2.571 & 1.258 & 0.984 & 1.031 & 8.623 \\
\textbf{slope} & -0.001 & 0.000 & 0.000 & 0.001 & 0.000 & 0.001 \\
\textbf{intercept} & 40.000 & 0.348 & 2.286 & 1.627 & 0.100 & 0.254 \\
\textbf{autocorrelation} & -0.115 & 0.707 & 0.982 & 0.805 & 0.787 & 0.378 \\
\textbf{self\_similarity} & 0.367 & 0.811 & 1.121 & 0.922 & 1.044 & 0.494 \\
\textbf{chaos} & 0.023 & 0.003 & 0.010 & 0.001 & 0.011 & -0.001 \\ \hline
\end{tabular}%
}
\resizebox{\textwidth}{!}{%
\begin{tabular}{rrrrrrr}
\hline
\textbf{Characteristic} & \textbf{\begin{tabular}[c]{@{}l@{}}Peer-\\ coin\end{tabular}} & \textbf{\begin{tabular}[c]{@{}l@{}}Vert-\\ coin\end{tabular}} & \textbf{\begin{tabular}[c]{@{}l@{}}Redd-\\ coin\end{tabular}} & \textbf{\begin{tabular}[c]{@{}l@{}}Feather-\\ coin\end{tabular}} & \textbf{\begin{tabular}[c]{@{}l@{}}Black-\\ coin\end{tabular}} & \textbf{\begin{tabular}[c]{@{}l@{}}Nova-\\ coin\end{tabular}} \\ \hline
\textbf{mean} & 10.085 & 2.502 & 4.646 & 2.005 & 1.090 & 6.819 \\
\textbf{standard\_deviation} & 47.070 & 0.180 & 68.175 & 6.443 & 0.105 & 2.295 \\
\textbf{skewness} & 30.324 & -1.782 & 20.761 & 11.521 & -4.368 & 24.326 \\
\textbf{kurtosis} & 919.356 & 30.015 & 434.280 & 157.793 & 18.525 & 891.281 \\
\textbf{maximum} & 1440.000 & 4.079 & 1440.000 & 130.909 & 1.335 & 96.000 \\
\textbf{minimum} & 1.377 & 0.151 & 0.646 & 0.148 & 0.442 & 0.451 \\
\textbf{lowerquant} & 7.742 & 2.412 & 0.986 & 1.042 & 1.111 & 6.154 \\
\textbf{median} & 8.372 & 2.500 & 1.007 & 1.048 & 1.114 & 6.606 \\
\textbf{upperquant} & 9.057 & 2.590 & 1.028 & 1.171 & 1.117 & 7.164 \\
\textbf{VaR99} & 5.464 & 2.144 & 0.935 & 1.034 & 0.551 & 4.364 \\
\textbf{VaR95} & 6.545 & 2.289 & 0.957 & 1.036 & 0.949 & 5.390 \\
\textbf{slope} & -0.003 & 0.000 & -0.010 & -0.002 & 0.000 & -0.001 \\
\textbf{intercept} & 1440.000 & 0.151 & 1440.000 & 0.291 & 1.309 & 1.765 \\
\textbf{autocorrelation} & 0.667 & 0.154 & 0.821 & 0.914 & 0.976 & 0.373 \\
\textbf{self\_similarity} & 0.717 & 0.437 & 1.051 & 1.210 & 1.337 & 0.697 \\
\textbf{chaos} & 0.002 & 0.008 & -0.001 & 0.032 & 0.006 & 0.009 \\ \hline
\end{tabular}%
}
\end{table}

\begin{table}[H]
\footnotesize
\singlespacing
\caption{Characteristics of Block size of different cryptocurrencies}
\label{tab:size_char}
\resizebox{\textwidth}{!}{%
\begin{tabular}{rrrrrrr}
\hline
\textbf{Characteristic} & \textbf{Bitcoin} & \textbf{Ethereum} & \textbf{Litecoin} & \textbf{\begin{tabular}[c]{@{}l@{}}Bitcoin\\ Cash\end{tabular}} & \textbf{\begin{tabular}[c]{@{}l@{}}Ethereum\\ Classic\end{tabular}} & \textbf{XRP} \\ \hline
\textbf{mean} & 407162.152 & 14376.916 & 12909.684 & 138173.724 & 1297.638 & NA \\
\textbf{standard\_deviation} & 363245.372 & 11337.562 & 15590.195 & 284058.956 & 340.581 & NA \\
\textbf{skewness} & 0.241 & 0.285 & 4.309 & 9.176 & 0.679 & NA \\
\textbf{kurtosis} & -1.583 & -0.819 & 31.780 & 109.791 & 2.106 & NA \\
\textbf{maximum} & 998092.000 & 58953.000 & 206020.000 & 4710539.000 & 3594.000 & NA \\
\textbf{minimum} & 134.000 & 575.164 & 134.000 & 4982.000 & 575.164 & NA \\
\textbf{lowerquant} & 21246.000 & 1627.750 & 4004.750 & 60455.500 & 1054.750 & NA \\
\textbf{median} & 310990.000 & 17024.000 & 7016.000 & 94775.000 & 1310.500 & NA \\
\textbf{upperquant} & 777369.500 & 23068.750 & 19366.500 & 122827.500 & 1492.250 & NA \\
\textbf{VaR99} & 134.548 & 658.423 & 561.630 & 15574.520 & 653.404 & NA \\
\textbf{VaR95} & 134.952 & 788.678 & 800.306 & 27169.700 & 775.052 & NA \\
\textbf{slope} & 266.541 & 17.464 & 8.806 & -89.253 & 0.189 & NA \\
\textbf{intercept} & 204.000 & 643.886 & 199.000 & 385996.000 & 643.886 & NA \\
\textbf{autocorrelation} & 0.985 & 0.981 & 0.872 & 0.626 & 0.850 & NA \\
\textbf{self\_similarity} & 1.067 & 1.310 & 1.148 & 1.074 & 1.131 & NA \\
\textbf{chaos} & 0.058 & 0.058 & 0.065 & 0.027 & 0.045 & NA \\ \hline
\end{tabular}%
}
\resizebox{\textwidth}{!}{%
\begin{tabular}{rrrrrrr}
\hline
\textbf{Characteristic} & \textbf{\begin{tabular}[c]{@{}l@{}}Bitcoin\\ SV\end{tabular}} & \textbf{Dash} & \textbf{Zcash} & \textbf{Monero} & \textbf{Dogecoin} & \textbf{\begin{tabular}[c]{@{}l@{}}Bitcoin \\ Gold\end{tabular}} \\ \hline
\textbf{mean} & 1100149.254 & 12999.389 & 23802.102 & 39874.397 & 10523.242 & 25312.953 \\
\textbf{standard\_deviation} & 1278250.457 & 26340.294 & 38911.209 & 47310.430 & 6607.125 & 67527.275 \\
\textbf{skewness} & 6.673 & 27.654 & 8.711 & 1.703 & 5.917 & 6.269 \\
\textbf{kurtosis} & 84.455 & 1040.743 & 117.847 & 4.063 & 68.981 & 45.828 \\
\textbf{maximum} & 20460199.000 & 1059232.000 & 687685.000 & 347816.000 & 116605.000 & 739259.000 \\
\textbf{minimum} & 5005.000 & 226.545 & 379.573 & 375.434 & 143.000 & 133.000 \\
\textbf{lowerquant} & 257789.500 & 3038.000 & 7189.500 & 3047.250 & 6775.000 & 6512.500 \\
\textbf{median} & 996071.500 & 9240.000 & 11670.000 & 20980.000 & 9510.000 & 9316.000 \\
\textbf{upperquant} & 1573243.000 & 19193.000 & 28242.000 & 62002.000 & 12022.000 & 14118.000 \\
\textbf{VaR99} & 6435.000 & 1312.960 & 2605.530 & 1058.990 & 3432.400 & 2727.870 \\
\textbf{VaR95} & 14660.750 & 1736.200 & 3103.900 & 1320.350 & 4491.000 & 3983.600 \\
\textbf{slope} & 2318.003 & 14.357 & -25.267 & 26.939 & 1.018 & -67.625 \\
\textbf{intercept} & 10871172.000 & 226.545 & 379.573 & 375.434 & 143.000 & 133.000 \\
\textbf{autocorrelation} & 0.377 & 0.298 & 0.836 & 0.958 & 0.798 & 0.618 \\
\textbf{self\_similarity} & 1.004 & 0.947 & 1.138 & 1.214 & 1.070 & 1.015 \\
\textbf{chaos} & 0.009 & 0.018 & 0.030 & 0.041 & 0.021 & -0.012 \\ \hline
\end{tabular}%
}
\resizebox{\textwidth}{!}{%
\begin{tabular}{rrrrrrr}
\hline
\textbf{Characteristic} & \textbf{\begin{tabular}[c]{@{}l@{}}Peer-\\ coin\end{tabular}} & \textbf{\begin{tabular}[c]{@{}l@{}}Vert-\\ coin\end{tabular}} & \textbf{\begin{tabular}[c]{@{}l@{}}Redd-\\ coin\end{tabular}} & \textbf{\begin{tabular}[c]{@{}l@{}}Feather-\\ coin\end{tabular}} & \textbf{\begin{tabular}[c]{@{}l@{}}Black-\\ coin\end{tabular}} & \textbf{\begin{tabular}[c]{@{}l@{}}Nova-\\ coin\end{tabular}} \\ \hline
\textbf{mean} & NA & 2641.881 & 772.025 & 806.556 & 687.622 & 539.712 \\
\textbf{standard\_deviation} & NA & 3611.409 & 634.442 & 1621.154 & 3441.373 & 1223.175 \\
\textbf{skewness} & NA & 3.420 & 3.613 & 10.605 & 28.388 & 38.218 \\
\textbf{kurtosis} & NA & 16.189 & 21.857 & 158.924 & 894.526 & 1712.453 \\
\textbf{maximum} & NA & 36709.000 & 7808.000 & 36789.000 & 120169.000 & 57527.000 \\
\textbf{minimum} & NA & 105.000 & 105.000 & 109.625 & 252.514 & 110.835 \\
\textbf{lowerquant} & NA & 682.104 & 388.361 & 359.746 & 286.296 & 360.352 \\
\textbf{median} & NA & 1149.000 & 526.043 & 460.827 & 386.251 & 436.181 \\
\textbf{upperquant} & NA & 3185.000 & 937.696 & 598.841 & 627.727 & 542.228 \\
\textbf{VaR99} & NA & 248.950 & 317.797 & 126.333 & 255.520 & 262.588 \\
\textbf{VaR95} & NA & 310.697 & 337.320 & 247.907 & 261.297 & 284.524 \\
\textbf{slope} & NA & -0.586 & -0.475 & -0.739 & -0.025 & -0.204 \\
\textbf{intercept} & NA & 130.000 & 175.000 & 109.625 & 464.500 & 141.000 \\
\textbf{autocorrelation} & NA & 0.894 & 0.609 & 0.705 & 0.360 & 0.069 \\
\textbf{self\_similarity} & NA & 1.129 & 1.007 & 1.063 & 0.959 & 0.951 \\
\textbf{chaos} & NA & 0.100 & 0.034 & 0.034 & 0.039 & 0.011 \\ \hline
\end{tabular}%
}
\end{table}

\FloatBarrier
\includepdf[pages=-]{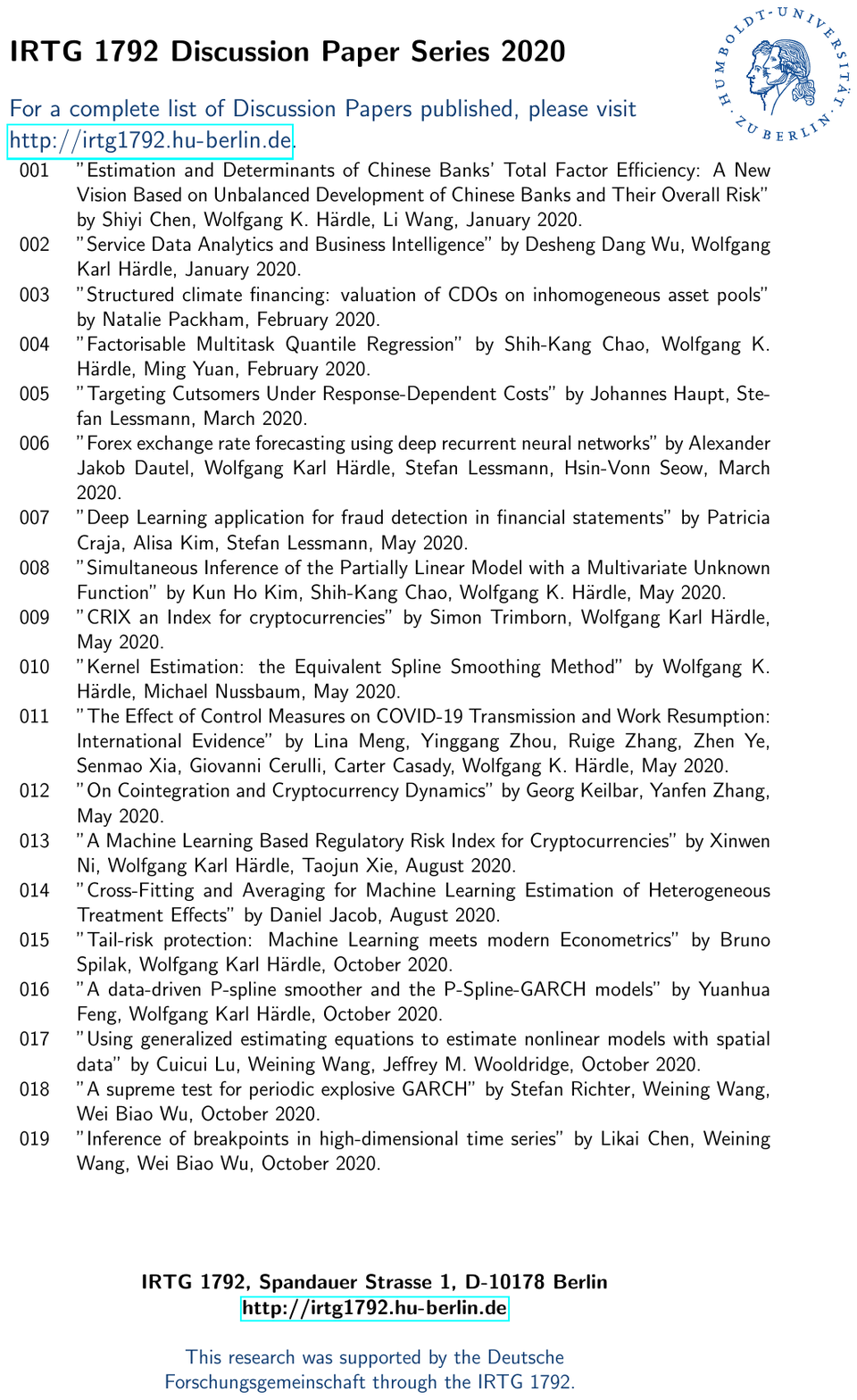}
\includepdf[pages=-]{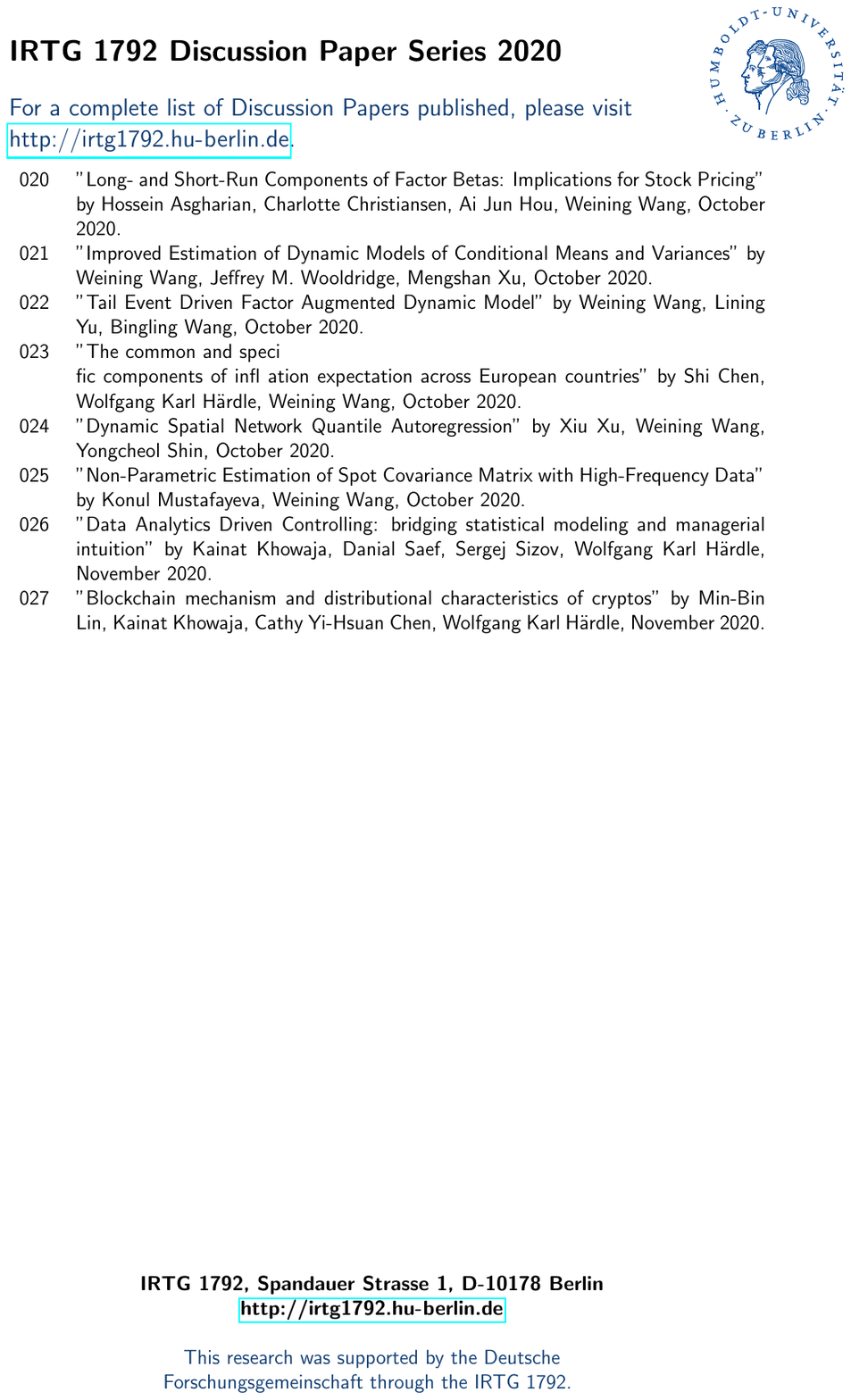}
\end{document}